# Observation of the Nernst effect driven by longitudinal spin fluctuations

Fuyuki Ando[1*], Hiroto Adachi[2], Hossein Sepehri-Amin[1], Takamasa Hirai[1,3], Keisuke Hirata[1,3], and Ken-ichi Uchida[1,3]

[1] *National Institute for Materials Science, Tsukuba, Japan*

[2] *Research Institute for Interdisciplinary Science, Okayama University, Okayama, Japan*

[3] *Department of Advanced Materials Science, Graduate School of Frontier Sciences, The University of Tokyo, Kashiwa, Japan*

[*] e-mail: ANDO.Fuyuki@nims.go.jp, ando1227fu@gmail.com

**The Nernst effect, which converts a heat current into a charge current in the orthogonal direction, is generally classified into the ordinary effect due to an external magnetic field applied to conductors and the anomalous effect due to static magnetization in magnetic materials. Thus, one expects that the anomalous Nernst effect disappears at the Curie temperature following the magnetization. Here, we observe the Nernst effect which rather manifests around the Curie temperature. In a series of ferromagnetic (Mn,Cr)Sb samples, the Nernst coefficients exhibit finite values even across the Curie temperatures, whereas the anomalous Hall resistivity disappears, suggesting the breakdown of the Mott relation. These surviving Nernst and vanishing Hall effects result in a peak behavior of the transverse thermoelectric conductivity around the Curie temperature, which we theoretically reproduce by introducing longitudinal fluctuations of spins without accounting for an exotic spin texture, a non-trivial electronic structure, or magnon- and phonon-drag effects. The Nernst effect driven by the longitudinal spin fluctuations reveals that not only the static magnetic order but also the spin fluctuations can be a driving force of the Nernst effect in magnetic materials, opening another way for boosting transverse thermoelectric conversion.**

The transverse thermoelectric conversion is the interconversion between heat and charge currents in the orthogonal direction. This orthogonal geometry simplifies the thermoelectric device architecture, eliminating additional components, such as substrates, electrodes, and their junctions, which leads to the improvement of thermoelectric conversion efficiency, thermal durability, and the reduction of mass production cost for thermoelectric modules[1–4]. To realize the transverse thermoelectric conversion, various mechanisms have been demonstrated related to magnetism or spin: the Nernst effect[5–17], spin Seebeck effect[18], and Seebeck-effect-driven Hall effects[19,20]; and those unrelated to magnetism or spin: the off-diagonal Seebeck effects[3,21–30] in natural anisotropic crystals and artificial anisotropic composites. The Nernst effect is generally classified into the ordinary and anomalous Nernst effects (ONE and ANE) due to an external magnetic field **H** applied to conductors and magnetization in magnets, respectively. Recently, ONE and ANE have been intensively studied associated with the emergence of the field of spin caloritronics and the development of topological materials science, such as Dirac semimetals and Weyl magnets[7–9,11,12,15].

Let us see the general understanding of ONE and ANE. As illustrated in Fig. 1, a transverse electric field $\mathbf{E}_{yx}$ generated by ONE and ANE in magnetic materials in the open circuit condition is expressed as,

$$\begin{aligned} -\mathbf{E}_{yx} &= S_{xy}\nabla_x T \\ &= S_{\mathrm{ONE}}(\mathbf{H}_z \times \nabla_x T) + S_{\mathrm{ANE}}(\mathbf{m}_z \times \nabla_x T), \end{aligned} \tag{1}$$

where $S_{xy}$ is the transverse thermopower, $S_{\mathrm{ONE}}$ and $S_{\mathrm{ANE}}$ are respectively the ordinary and anomalous Nernst coefficients, $\nabla_x T$ is the temperature gradient along the $x$ direction, and $\mathbf{m}_z$ is the unit magnetization vector along the $z$ direction. ONE is the Lorentz-force-induced deflection of charge carriers and $S_{\mathrm{ONE}}$ is determined by the energy dependence of carrier mobility regardless of the magnetic phase[5]. Meanwhile, $S_{\mathrm{ANE}}$ is phenomenologically separated into $S_{\mathrm{I}}$ and $S_{\mathrm{II}}$ as,

$$\begin{aligned} S_{\mathrm{ANE}} &= S_{\mathrm{I}} + S_{\mathrm{II}} \\ &= \rho_{xx}\alpha_{xy} - \rho_{\mathrm{AHE}}\alpha_{xx}, \end{aligned} \tag{2}$$

where $\rho_{xx}$ is the longitudinal electrical resistivity, $\rho_{\mathrm{AHE}}$ is the transverse electrical resistivity due to the anomalous Hall effect (AHE), $\alpha_{xx}$ and $\alpha_{xy}$ are respectively the diagonal and off-diagonal components of the Peltier tensor[7]. To date, $\alpha_{xy}$ has been understood to follow the Mott relation with AHE[7–9,11,12,15,31,32], whose origin is classified into the intrinsic mechanism by the Berry curvature in a spin-split electronic band structure, extrinsic mechanisms such as skew scattering and side-jump, and magnon- and phonon-drag effects[14,33,34]. $S_{\mathrm{II}}$ is the hybrid action of AHE and the Seebeck effect. Taking all the principles into account, $S_{\mathrm{ANE}}$ in typical ferromagnets emerges only when the magnetic moments are ordered and hence is expected to disappear at the Curie temperature $T_{\mathrm{C}}$.

Here, we propose the Nernst effect driven by the longitudinal fluctuations of spins. Figure 1 shows a schematic concept of the spin-fluctuation-driven Nernst effect (SFNE), whose contribution on $S_{xy}$ maximizes around $T_C$ contrary to the expectation from the conventional ANE. We characterized $S_{xy}$ for ferromagnetic $Mn_{1-a}Cr_aSb$ compounds with the strong *s*-*d* coupling allowing for the magnetic phase to influence the Seebeck coefficient $S_{xx}$ and $T_C$ systematically tuned by the doped Cr composition *a*. The temperature *T* dependence of $S_{xy}$ for $Mn_{1-a}Cr_aSb$ reveals that the finite $S_{xy}$ value remains across the ferromagnetic-paramagnetic phase transitions, whereas $\rho_{AHE}$ disappears. The surviving $S_{xy}$ and vanishing $\rho_{AHE}$ ($S_{II}$) results in a peak behavior of $\alpha_{xy}$ ($S_I$) around $T_C$, which is theoretically reproduced by considering longitudinal spin fluctuations and solving the Kubo formula. Our observation of SFNE reveals that not only the static magnetic order but also the fluctuations of spins can drive the Nernst effect, opening another way for boosting transverse thermoelectric conversion.

## Fundamental properties of (Mn,Cr)Sb

We synthesized the $Mn_{1-a}Cr_aSb$ polycrystals with $a$ = 0, 0.2, 0.4, 0.6 and characterized the crystal structures and magnetic properties. MnSb is a semimetal having a NiAs-type hexagonal crystal structure and itinerant ferromagnetism with $T_C$ much higher than room temperature, where the doped Cr forms a complete solid solution at the Mn site[35,36] (Fig. 2a drawn using the VESTA software package[37]). In this work, a series of $Mn_{1-a}Cr_aSb$ polycrystals was synthesized by induction melting followed by pulverization to powders. The powders were consolidated using a spark plasma sintering method (see Methods for details). Figure 2b shows the X-ray diffraction curves for the $Mn_{1-a}Cr_aSb$ powders, confirming that all the samples have a hexagonal structure without the other intermetallic phases. As shown in Extended Data Fig. 1, slight precipitations of pure Cr were observed in the Cr-doped samples because of its higher melting point than those of Mn and Sb. Figure 2c shows the magnetic field *H* dependence of the magnetization *M* at 300 K for the sintered $Mn_{1-a}Cr_aSb$ samples. The pure MnSb shows a ferromagnetic behavior whose *M* saturates around $\mu_0 H$ = 0.3 T, whereas the magnetization decreases as *a* increases. We estimated $T_C$ for the sintered $Mn_{1-a}Cr_aSb$ samples in Fig. 2d by measuring the *M*–*T* curves and applying the Arrott plot as shown in Extended Data Fig. 2. As a result, $T_C$ of the $Mn_{1-a}Cr_aSb$ samples with $a$ = 0, 0.2, 0.4, 0.6 systematically differs, being 480, 450, 385, and 295 K, respectively, whose trend agrees with the previous report[36].

## Characterization of temperature dependence of the Nernst coefficient

To characterize $S_{xy}$ for the $Mn_{1-a}Cr_aSb$ samples, we measured the Ettingshausen effect, the reciprocal effect of the Nernst effect, using a lock-in thermography (LIT) technique[13,38,39]. The reason why we adopted the LIT method is that it enables the precise quantification of $S_{xy}$ in isothermal condition, whereas the direct Nernst measurement for the bulk samples inevitably includes the parasitic contribution by the off-diagonal thermal conduction[40] due to the anisotropic thermal conductivity[41] or the thermal Hall effect[42,43]. Figure 3a illustrates the LIT measurement setup and extraction process of thermoelectric signals. As shown in a photograph, the bar-shaped $Mn_{1-a}Cr_aSb$ samples were aligned on an insulating substrate and formed an electrical series circuit to flow charge current **I** in a zigzag manner. Under the application of an in-plane **H** with $\mu_0 H$ = 0.33 T, a square-wave **I** with amplitude $I$ = 1 A orthogonal to **H** was applied to the samples. We can expect that the Ettingshausen effect generates $\nabla T$ in the cross product direction of **I** and **H**, and thus measured the resultant $T$ modulations of the top surfaces oscillating at the same frequency $f$ as **I** by an infrared camera. The obtained thermal images were transformed into lock-in amplitude $A$ and phase $\varphi$ images through the Fourier analysis, which enables the extraction of the thermoelectric (Peltier and Ettingshausen) effects from a Joule heating contribution because the Joule-heating-induced $T$ modulation is constant regardless of the sign of $I$ (Fig. 3a). In addition, to extract the pure Ettingshausen signals, which have an odd dependence on $H$, we obtained the LIT images under $\mu_0 H = \pm 0.33$ T and analyzed them into $A_{odd}$ and $\varphi_{odd}$ images according to Eqs. (5,6) in Methods. The magnitude and sign of $S_{xy}$ were determined from the $A_{odd}$ and $\varphi_{odd}$ values, respectively, as discussed later.

Figure 3b–d shows the $T$ dependence of the topography, $A_{odd}$, and $\varphi_{odd}$ images at $f$ = 1 Hz for the $Mn_{1-a}Cr_aSb$ samples, respectively. As shown in Fig. 3b, the top surfaces of all the samples were captured in a single frame and the base $T$ was controlled by a resistance heater in the sample holder. The surface of the sample with larger $a$ tends to show a higher $T$, reflecting the larger Joule heating due to $\rho_{xx}$ (see Extended Data Fig. 3b). In Fig. 3c,d, we observed the clear and homogeneous $A_{odd}$ and $\varphi_{odd}$ signals derived from the Ettingshausen effect on the sample surfaces. The $A_{odd}$ signals decrease as $a$ increases at the same base $T$ and become undetectable for the sample with $a$ = 0.6 ($T_C$ = 295 K) over the entire $T$ range. As base $T$ increases, the $A_{odd}$ signals decrease to zero above $T_C$ of each sample, suggesting the disappearance of $S_{xy}$. Meanwhile, for the samples with $a$ = 0 and 0.4 (0.2) where **I** flows in the down-to-up (up-to-down) direction, the $\varphi_{odd}$ signals are ~ 0° (~ 180°) suggesting the increase (decrease) of $T$ on the top surfaces. In comparison with the $\varphi_{odd}$ signals characterized by the same LIT method for the other ferromagnetic systems[10,13,17], the sign of $S_{xy}$ for the $Mn_{1-a}Cr_aSb$ samples with $a$ = 0, 0.2, 0.4 is found to be negative.

The magnitude of $S_{xy}$ needs to be determined at thermally steady state. Thus, we measured the $f$ dependence of $A_{odd}$ and $\varphi_{odd}$ to estimate the $A_{odd}$ value at steady state ($f$ = 0 Hz). Figure 3e,f respectively shows the $f$ dependence of the average $A_{odd}$ and $\varphi_{odd}$ signals within the marked area in Fig. 3b measured around room temperature. As $f$ increases, the magnitude of $A_{odd}$ gradually decreases and $\varphi_{odd}$ slightly deviates from 0° and 180° due to the delay of the $T$ modulation due to heat diffusion induced by the Ettingshausen effect. By solving the one-dimensional heat diffusion equation in $f$ domain as indicated by the solid curves in Fig. 3e, we reproduced the $f$ dependence of $A_{\mathrm{odd}}$ and obtained the $A_{\mathrm{odd}}$ values at the steady state ($f = 0$ Hz) by extrapolating the fitting curves.

Figure 3g shows the $T$ dependence of $A_{\mathrm{odd}}$ at $f = 0$ Hz ($A_{\mathrm{odd}}^{0\,\mathrm{Hz}}$) normalized by the applied charge current density $j$ for all the $Mn_{1-a}Cr_aSb$ samples. Here, $T$ is directly determined from the surface $T$ for each sample in topography images (Fig. 3b). For the $Mn_{1-a}Cr_aSb$ samples with $a = 0, 0.2, 0.4$, $A_{\mathrm{odd}}^{0\,\mathrm{Hz}}/j$ shows finite values from room temperature to $T_{\mathrm{C}}$ and gradually decreases to zero above 500 K. The sample with $a = 0.6$ shows almost zero values over the entire $T$ range. These results suggest that $S_{\mathrm{ONE}}$ is negligibly small compared with $S_{\mathrm{ANE}}$ for the $Mn_{1-a}Cr_aSb$ samples, which is consistent with the previous reports on (Mn,Cr)Sb thin films[44,45]. Thus, we analyze $S_{xy}$ under the assumption of $S_{xy} \approx S_{\mathrm{ANE}}$ in Eqs. (1,2) in the following discussion.

**Observation of SFNE around Curie temperature**

Now, we are in a position to characterize the $T$ dependence of $S_{xy}$, $S_{\mathrm{I}}$, and $S_{\mathrm{II}}$ according to Eq. (2) and compare them with the spontaneous magnetization $M_{\mathrm{s}}$. Figure 4a,b shows the results for the $Mn_{1-a}Cr_aSb$ samples with $a = 0$ and 0.2 having higher $T_{\mathrm{C}}$. $S_{xy}$ is analyzed from the LIT results (see Methods), $S_{\mathrm{I}}$ and $S_{\mathrm{II}}$ by measuring the $T$ dependence of $S_{xx}$, $\rho_{xx}$, and $\rho_{\mathrm{AHE}}$ as shown in Extended Data Figs. 3 and 4, and $M_{\mathrm{s}}$ from the intercept values for the Arrott plot in Extended Data Fig. 2. For both samples, the $T$ dependence of $S_{xy}$ does not follow that of $M_{\mathrm{s}}$, showing finite values across the ferromagnetic-paramagnetic phase transition. On the other hand, $\rho_{\mathrm{AHE}}$ (Extended Data Fig. 4) and thus $S_{\mathrm{II}}$ decrease to zero exactly at $T_{\mathrm{C}}$ following $M_{\mathrm{s}}$. This contradiction between $S_{xy}$ and $\rho_{\mathrm{AHE}}$ suggests the breakdown of the Mott relation[31,32] in the vicinity of $T_{\mathrm{C}}$. Importantly, as shown in the middle panel of Figs. 4a,b, the $S_{\mathrm{I}}$ term in Eq. (2) exhibits a peak-like behavior around $T_{\mathrm{C}}$ for both the samples. Because $\rho_{xx}$ moderately changes across the phase transition, this anomaly is attributed to $\alpha_{xy}$ in Eq. (2). Figure 4c shows $\alpha_{xy}$ as a function of $T$ normalized by $T_{\mathrm{C}}$ for the $Mn_{1-a}Cr_aSb$ samples with $a = 0$ and 0.2, which shows the peak near $T/T_{\mathrm{C}} = 1$ while $\alpha_{xy}$ derived from ANE is expected to vanish as well as $\rho_{\mathrm{AHE}}$.

How do we explain the origin of this anomaly in $\alpha_{xy}$ for the $Mn_{1-a}Cr_aSb$ samples? The first possible mechanism to explain this contradiction between $S_{xy}$ and $\rho_{\mathrm{AHE}}$ is the drag effect by quasiparticles such as magnons. Since quasiparticles carry heat but not electricity, they significantly contribute to thermoelectric conversion, while rarely contributing to electrical conduction. The magnon-drag effect has historically been studied as the relationship between spin dynamics and thermoelectrics mainly on the Seebeck effect[36,46–48], and recently reported on $\alpha_{xy}$ in ferromagnetic MnBi[14]. However, the magnon-drag contribution was naturally limited to much lower $T$ than $T_{\mathrm{C}}$ because of the decrease of the magnon lifetime in the high $T$ region. Thus, our results in the vicinity of $T_{\mathrm{C}}$ cannot be explained by the drag effect by quasiparticles. Second, the spin-fluctuation effect has been suggested to explain the anomaly on the Seebeck effect around $T_C$ in itinerant antiferromagnetic and weak ferromagnetic systems[49,50] and is expected to also influence the Nernst effect[32] . However, the $S_{II}$ term including $S_{xx}$ in our ferromagnetic system decreases to zero exactly at $T_C$, suggesting the absence of such Seebeck-effect-based contributions. Finally, although the anomalies on $\alpha_{xy}$ around the magnetic phase transition or the breakdown of the Mott relation were reported in several systems, the physical origins are the fluctuation of chiral spin texture or the anomaly of the Berry curvature[9,51,52], which also exhibit an anomaly in $\rho_{AHE}$ with the finite values. Thus, these mechanisms for $S_{ANE}$ proposed in previous works cannot explain the anomaly of $\alpha_{xy}$ due to the surviving $S_{xy}$ and vanishing $\rho_{AHE}$ observed in this work.

To theoretically explain our experimental results, we introduce the idea of longitudinal spin fluctuations (fluctuation of spins parallel to the spin quantization axis), whereas magnons and paramagnons are defined by transverse spin fluctuations (fluctuation of spins perpendicular to the spin quantizing axis). The effects of longitudinal spin fluctuations on the transport phenomena have long been studied since Fisher and Langer discussed the anomalies in electrical resistivity near the magnetic transition[53]. We apply the same concept to the Nernst effect in ferromagnets and calculate $\alpha_{xy}$ derived from SFNE ($\alpha_{xy}^{\mathrm{SFNE}}$), in a manner similar to the diagrammatic calculation of the vortex Nernst effect[54] (see Methods and Extended Data Fig. 5). The solid line in Fig. 4c is $\alpha_{xy}^{\mathrm{SFNE}}$ calculated from the Kubo formula, which shows a cups shape around $T_C$ and clearly reproduces our experimental results of $\alpha_{xy}$ especially in the paramagnetic regime. This agreement validates our observation of the surviving $S_{xy}$ and vanishing $\rho_{AHE}$ is explained by the contribution of SFNE. Meanwhile, the calculated $\alpha_{xy}^{\mathrm{SFNE}}$ and experimentally analyzed $\alpha_{xy}$ differ from each other in the ferromagnetic region probably because we phenomenologically introduce the coefficient $\Lambda_{xy}$ as a $T$-independent parameter [see Eq. (9) in Methods] for but should have the $T$ dependence according to the intrinsic and extrinsic scattering mechanisms.

**Prospects and summary**

Finally, let us note prospects on how to study SFNE toward the high-performance transverse thermoelectric conversion. For materials development, our proposal will pave a wider exploration space in terms of temperature and magnetic materials. Whereas the magnon-drag contribution on ANE is limited to low-$T$ regions, this work demonstrates the SFNE contribution far above room temperature, leading to versatile thermoelectric applications. From the theoretical point of view [see Eq. (10) in Methods], the exploration of magnetic materials with the larger $\Lambda_{xy}$ and $\Gamma_2$ values is necessary. $\Lambda_{xy}$ is the effective anomalous velocity in transverse direction induced by the spin-orbit interaction and $\Gamma_2$ can arise from microscopic considerations taking account of particle-hole asymmetry. Meanwhile, because our proposed mechanism is independent of the other transverse thermoelectric effects, the idea of hybrid transverse magneto-thermoelectric conversion can include the SFNE without any conflicts [26,27,29,30].

In conclusion, we demonstrate the Nernst effect driven by the longitudinal spin fluctuations. In the ferromagnetic (Mn,Cr)Sb samples with different $T_C$, $S_{xy}$ exhibits finite values across $T_C$ while $\rho_{AHE}$ disappears, suggesting the broken Mott relation. $\alpha_{xy}$ shows a peak around $T_C$, which is theoretically reproduced by introducing the longitudinal fluctuations of spins without accounting for an exotic spin texture, a non-trivial electronic structure, or the magnon-drag effect. This work manifests that not only the static magnetic order but also the spin fluctuations can be a driving force of the Nernst effect in magnetic materials, opening another way for boosting transverse thermoelectric conversion.

**Methods**

**Sample preparation.** $Mn_{1-a}Cr_aSb$ ($a$ = 0, 0.2, 0.4, 0.6) alloys were fabricated through induction melting and casting into a steel die of high-purity elements (>99.9%), followed by annealing at 750℃ for 24 hours and grinding into powders. All the processes were conducted in an Ar atmosphere. The obtained powders were densified by a spark plasma sintering method at 700℃ for 20 min under a unidirectional pressure of 50 MPa in a vacuum. The relative density of the $Mn_{1-a}Cr_aSb$ samples is higher than 95%.

**Sample characterization.** Powder X-ray diffraction $2\theta$-$\theta$ curves (Fig. 2b) were obtained using an X-ray diffractometer with Cr-$k_\alpha$ radiation (MiniFlex, Rigaku Corp.). Elemental maps of the mechanically polished samples were observed by a scanning electron microscope with energy-dispersive X-ray spectroscopy (SEM-EDX) using Cross-Beam 1540ESB (Extended Data Fig. 1). The temperature $T$ and magnetic-field $H$ dependences of magnetization $M$ (Extended Data Fig. 2a) were measured by the superconducting quantum interference device vibrating sample magnetometry using Magnetic Properties Measurement System with an oven option (MPMS3, Quantum Design Inc.). The Arrott plot was applied as shown in Extended Data Fig. 2b to estimate the spontaneous magnetization $M_s$ from the intercept and Curie temperature $T_C$ from the $T$ dependence of $M_s$ (Extended Data Fig. 2c).

**Ettingshausen measurement using a lock-in thermography method.** The samples with a dimension of approximately 8 mm × 2 mm × 1 mm were fixed on a sapphire substrate with Aron Ceramic D-type (Toagosei Co., Ltd.) and attached to a Cu stage (Fig. 3a), where the base temperature can be controlled by an embedded heater and resistance temperature sensor. To enhance the infrared emissivity and ensure uniform emission properties, the top surface of the samples was coated with an insulating black ink having an emissivity higher than 0.94 (JSC-3, JAPANSENSOR Corp.). The LIT measurements were performed under the application of an in-plane magnetic field of 0.33 T in a high vacuum and thermal images were monitored through an infrared-transparent $CaF_2$ window. The relationship between the obtained infrared intensity and absolute temperature influenced by the black ink and $CaF_2$ window was calibrated beforehand by measuring the infrared intensity on the surface of a reference thermometer under the same condition, which allows us to precisely determine the sample temperature during the LIT measurements through topography images (Fig. 3b). A rectangularly-modulated alternating charge current with the amplitude 1 A, frequencies $f$ (= 1.0, 2.0, 4.0, 5.0, 10.0 Hz), and zero offset was applied to the samples perpendicular to the magnetic field direction. Then, the first harmonic responses of the spatial distribution of infrared radiation thermally emitted from the sample surface were extracted and transformed into amplitude $A$ and phase $\varphi$ images through the Fourier analysis. These LIT images were obtained under ±0.33 T and separated into the $H$-even- and $H$-odd-dependent components using the following equations:

$$A_{\text{even}} = \left|A(+H)e^{-i\varphi(+H)} + A(-H)e^{-i\varphi(-H)}\right|/2, \tag{3}$$

$$\varphi_{\text{even}} = -\arg\left[A(+H)e^{-i\varphi(+H)} + A(-H)e^{-i\varphi(-H)}\right], \tag{4}$$

$$A_{\text{odd}} = \left|A(+H)e^{-i\varphi(+H)} - A(-H)e^{-i\varphi(-H)}\right|/2, \tag{5}$$

$$\varphi_{\text{odd}} = -\arg\left[A(+H)e^{-i\varphi(+H)} - A(-H)e^{-i\varphi(-H)}\right], \tag{6}$$

where $A_{\text{even}}$ ($A_{\text{odd}}$) and $\varphi_{\text{even}}$ ($\varphi_{\text{odd}}$) represent $A$ and $\varphi$ components having the $H$-even ($H$-odd) dependence, respectively. $A_{\text{odd}}$ and $\varphi_{\text{odd}}$ in Fig. 3c,d reflect the pure Ettingshausen contribution.

**Transport measurements.** Thermoelectric transport properties were characterized as follows. The $T$ dependence of the Seebeck coefficient $S_{xx}$ and electrical resistivity $\rho_{xx}$ shown in Extended Data Fig. 3a,b was measured using the Seebeck-coefficient/electric-resistance measurement system (ZEM-3, ADVANCE RIKO Inc.). The $T$ dependence of the thermal conductivity $\kappa_{xx}$ shown in Extended Data Fig. 3c was determined through thermal diffusivity measured using the laser flash method, specific heat measured using the differential scanning calorimetry, and density. The off-diagonal component of the resistivity tensor $\rho_{xy}$ shown in Extended Data Fig. 4 was measured using a Physical Property Measurement System (PPMS, Quantum Design Inc.). To increase $T$ above 400 K, a home-made sample pack embedded with a ceramic heater and T-type thermocouple was used. Rectangular-shaped samples with a dimension of approximately 10 mm × 2 mm × 1 mm were set on the ceramic stage with Aron Ceramic D-type (Toagosei Co., Ltd.). A dc charge current of 100 mA was applied along the length direction, and the transverse Hall voltage was measured across the width direction. $H$ was applied perpendicular to the current (out-of-plane) and swept between −3 and + 3 T. Extended Data Fig. 4a,b plots $\rho_{xy_\text{odd}} = [\rho_{xy}(+H)- \rho_{xy}(-H)]/2$ for the $Mn_{1-a}Cr_aSb$ ($a$ = 0 and 0.2) samples. Extended Data Fig. 4c plots $\rho_{\text{AHE}}$ defined as the intercept value of linear fitting for $\rho_{xy_\text{odd}}$ in the range of 2–3 T.

**Semi-phenomenological modeling of SFNE.** To calculate the transverse thermoelectric conductivity by SFNE $\alpha_{xy}^{\text{SFNE}}$, we consider the following Gaussian action for the longitudinal spin fluctuations[55] (in the unit $\hbar = k_{\text{B}} = c = 1$ with $\hbar$, $k_{\text{B}}$, and $c$ respectively being the Planck constant divided by $2\pi$, Boltzmann constant, and the velocity of light),

$$A = \sum_{\mathbf{q},\omega_n} \delta m^*_{\mathbf{q}}(\omega_m)\, D^{-1}_{\mathbf{q}}\,(i\omega_m)\, \delta m_{\mathbf{q}}(\omega_m) + J_{sd} \sum_{\mathbf{q},\omega_m} \delta m^*_{\mathbf{q}}(\omega_m)\; \sigma^z_{\mathbf{q}}(\omega_m), \tag{7}$$

where $\delta m^*_{\mathbf{q}}(\omega_m)$ is the longitudinal component of spin fluctuations with momentum $\mathbf{q}$ (with the magnitude of $q$) and Matsubara frequency $\omega_m = 2\pi T m$, $D^{-1}_{\mathbf{q}}(i\omega_m) = \epsilon_0 v_0[\mu_{GL} + \xi_0^2 q^2 + |\omega_m|/\Gamma]$ is the inverse of fluctuation propagator with $\epsilon_0$ and $v_0$ being the magnetic energy density and the volume of magnetic unit cell, respectively, $\xi_0$ being the bare correlation length, and $\Gamma$ is the damping constant for a spin non-conserving system. Moreover, $J_{sd}$ is the *s*-*d* exchange coupling constant, $\sigma^z_{\mathbf{q}}(\omega_m) = \sqrt{T/N}\sum_{\mathbf{p},\varepsilon_n} c^\dagger_{\mathbf{p}}(\varepsilon_n)\, \hat{\sigma}_z c_{\mathbf{p+q}}(\varepsilon_n + \omega_m)$ is the $z$ component of electron spin density for electron field variable $c_{\mathbf{p}}(\varepsilon_n)$ with momentum $\boldsymbol{p}$ and Matsubara frequency $\varepsilon_n = 2\pi T(n + \frac{1}{2})$, and $N$ is the total number of lattice sites. Note that the above action is derived from the Ginzburg-Landau action by an expansion about the mean-field solution, such that the mass term $\mu_{\mathrm{GL}}$ is given by $\mu_{\mathrm{GL}} = |a_{\mathrm{GL}}| + b_{\mathrm{GL}} m^2_{eq}$, where $a_{\mathrm{GL}} = (T - T_{\mathrm{C}})/T_{\mathrm{C}}$ with $T_{\mathrm{C}}$ being the bare Curie temperature, $b_{\mathrm{GL}}$ is the quartic coefficient of the Ginzburg-Landau action, and $m^2_{eq} = |a_{\mathrm{GL}}|/\, b_{\mathrm{GL}}$ below $T_{\mathrm{C}}$ and $m^2_{eq} = 0$ above $T_{\mathrm{C}}$. Note also that there is no anomaly near $T_{\mathrm{C}}$ in the transverse spin fluctuations (magnons), such that it is discarded here.

We compute $\alpha^{\mathrm{SFNE}}_{xy}$ using the Kubo formula[52]. Considering the longitudinal spin-fluctuation-driven process shown in Extended Data Fig. 5a and performing the analytic continuation of Matsubara frequency[56], we obtain

$$\alpha^{\mathrm{SFNE}}_{xy} = -\frac{|e|}{2T^2}\int \frac{d^3q}{(2\pi)^3} \int_{-\infty}^{\infty} \frac{d\omega}{2\pi} V^{\mathrm{SFNE}}_y(\boldsymbol{q}) v_x(\boldsymbol{q}) \frac{\omega}{\sinh^2\left(\frac{\omega}{2T}\right)} \left(\mathrm{Im}\, D^R_{\boldsymbol{q}}(\omega)\right)^2, \tag{8}$$

where $D^R_{\mathbf{q}}(\omega) = D_{\mathbf{q}}(i\omega_m \to \omega + i0^+)$ is the retarded fluctuation propagator. Note that $V^{\mathrm{SFNE}}_y(\mathbf{q})$ comes from the hatched triangle in Extended Data Fig. 5a, which is the charge current vertex renormalized by the spin-orbit interaction and the *s*-*d* exchange interaction. Intuitively, $V^{\mathrm{SFNE}}_y(\mathbf{q})$ can be regarded as an anomalous velocity induced by the spin-orbit interaction [e.g., see Eq. (3.5) in the literature[57]]. Guided by this intuitive picture, we set

$$V^{\mathrm{SFNE}}_y(\mathbf{q}) = -\Lambda_{xy} q_x \tag{9}$$

by introducing a phenomenological constant $\Lambda_{xy}$. Then, in performing the frequency integral of Eq. (8), we assume that the damping coefficient has an imaginary part $\Gamma = \Gamma_1 + i\Gamma_2$ as was done in the context of superconducting fluctuations under magnetic field[58]. Now, after doing the frequency integral as well as performing the momentum integral, $\alpha^{\mathrm{SFNE}}_{xy}$ is calculated to be

$$\alpha_{xy}^{\mathrm{SFNE}} = \frac{|e|\Lambda_{xy}\Gamma_2}{12\pi^2\xi_0^2\epsilon_0 v_0} F(Q) \tag{10}$$

where $Q = \xi_0 q_{cut}$ with the momentum cutoff $q_{cut}$, and

$$F(Q) = Q + \frac{\mu_{\mathrm{GL}} Q}{2(Q^2 + \mu_{\mathrm{GL}})} - \frac{3\sqrt{\mu_{\mathrm{GL}}}}{2} \tan^{-1}\left(\frac{Q}{\sqrt{\mu_{\mathrm{GL}}}}\right). \tag{11}$$

In Fig. 4c, we set $Q = 0.5$ and $|e|\Lambda_{xy}\Gamma_2/12\pi^2\xi_0^2\epsilon_0 v_0 = -0.025$ for $\alpha_{xy}^{\mathrm{SFNE}}$ to qualitatively discuss the origin of $\alpha_{xy}$ around $T_{\mathrm{C}}$.

**Data availability**

The data that support the findings of this study are available from the corresponding authors upon reasonable request.

**Acknowledgements**

The authors thank W. Zhou and Y. Sakuraba for technical support and valuable discussions. This work was supported by ERATO "Magnetic Thermal Management Materials" (No. JPMJER2201) from Japan Science and Technology Agency (JST), Grants-in-Aid for Scientific Research (KAKENHI) (No. 24K17610 and No. 26K00658) from Japan Society for the Promotion of Science (JSPS), and Iketani Science and Technology Foundation.

**Author contributions**

F.A. conceived the idea and supervised the study. H.S. and F.A. synthesized the polycrystalline samples. H.S. collected the XRD and SEM-EDX data. F.A. performed the LIT measurements with the help of T.H. and K.H. F.A. collected and analyzed the data on magnetic and thermoelectric properties, and LIT images. H.A. developed the theoretical model. F.A. and K.U. prepared the manuscript and all the authors discussed the results and commented on the manuscript.

**Competing interests**

The authors have no conflicts to disclose.

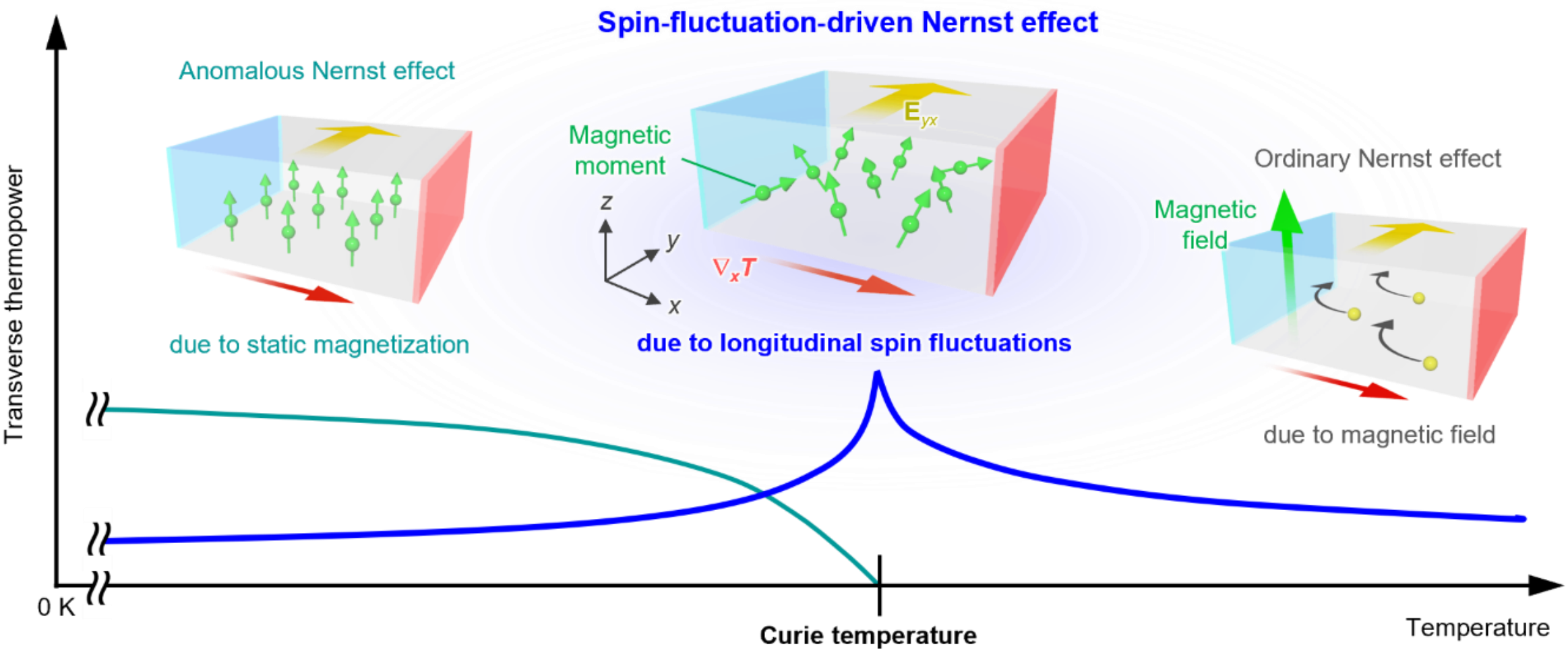

**Fig. 1 | Concept of the spin-fluctuation-driven Nernst effect.** Schematic of the spin-fluctuation-driven Nernst effect (SFNE) and the ordinary and anomalous Nernst effects (ONE and ANE) with their typical spin texture and temperature dependence around the Curie temperature $T_C$. The enhanced longitudinal spin fluctuations drive SFNE, which maximizes around $T_C$ contrary to the expectation from ANE.

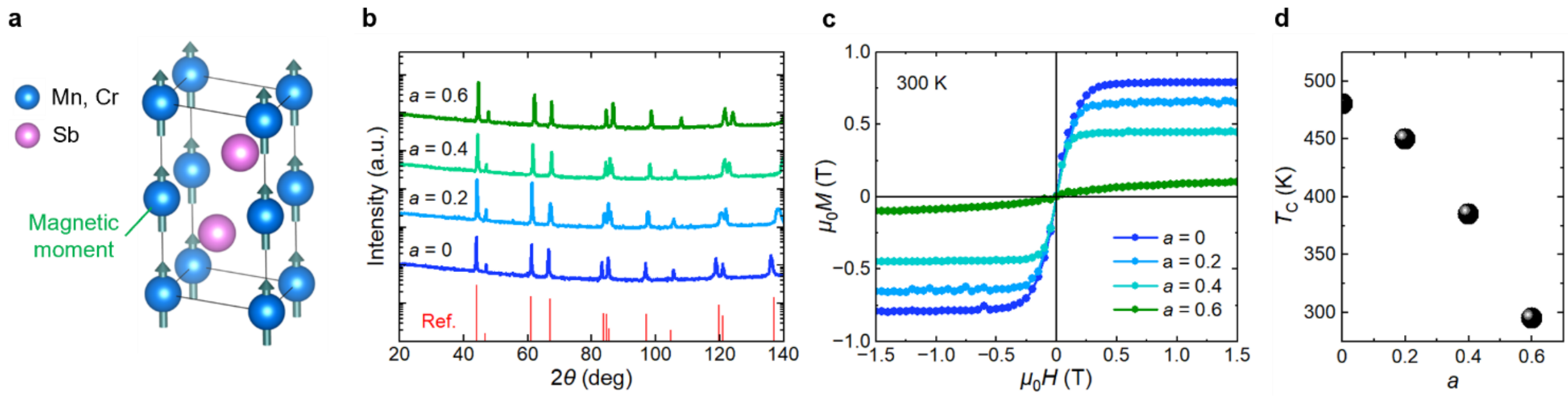

**Fig. 2 | Crystal structure and magnetic property for $Mn_{1-a}Cr_aSb$. a**, NiAs-type hexagonal crystal structure and itinerant ferromagnetism of MnSb. **b**, X-ray diffraction $2\theta$-$\theta$ curves for the $Mn_{1-a}Cr_aSb$ powders with the reference pattern of hexagonal MnSb[35] . **c**, The magnetic field $H$ dependence of the magnetization $M$ at 300 K for the sintered $Mn_{1-a}Cr_aSb$ samples. **d**, $T_C$ for the sintered $Mn_{1-a}Cr_aSb$ samples as a function of $a$.

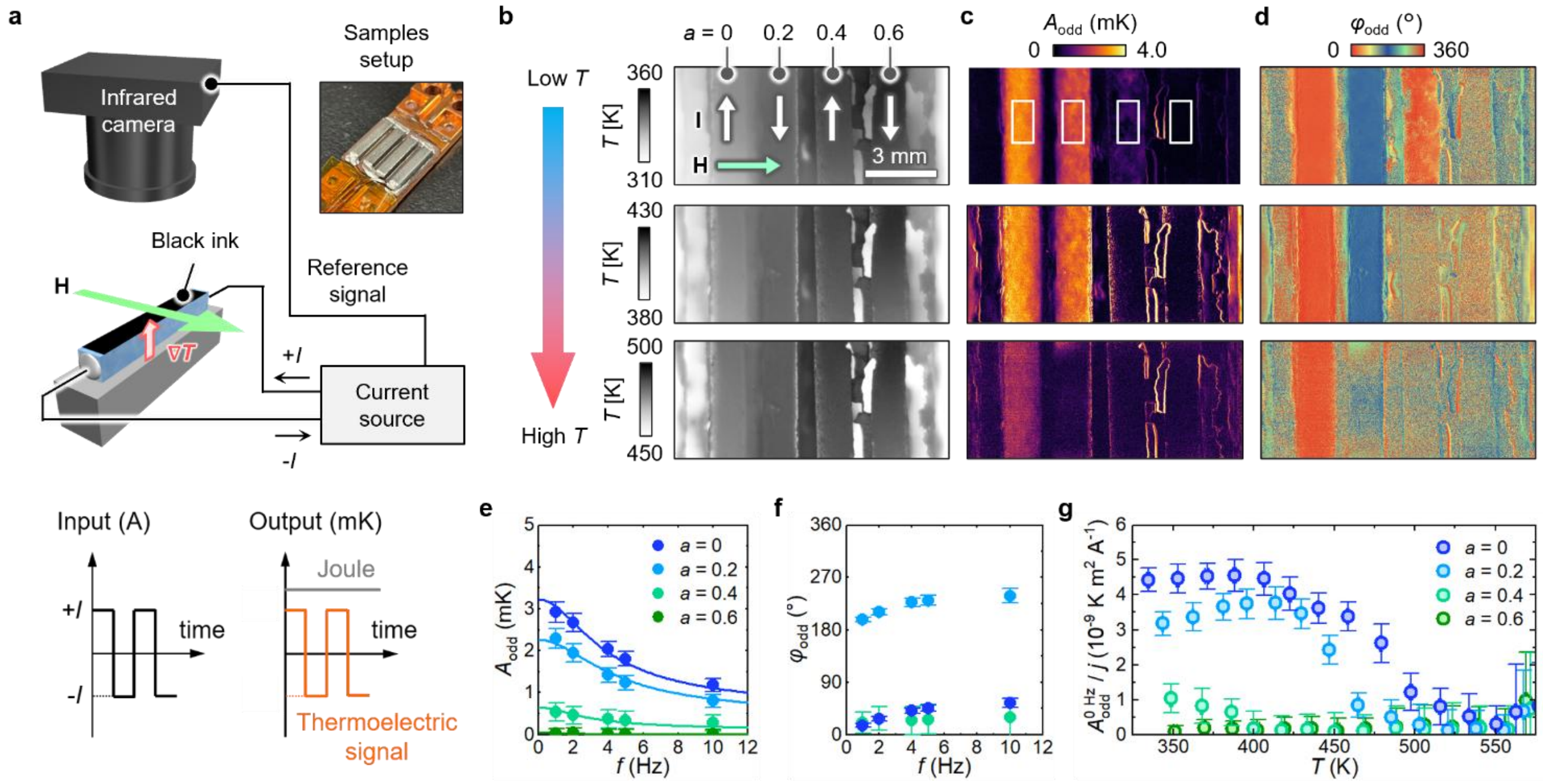

**Fig. 3 | Ettingshausen measurement based on a lock-in thermography technique. a**, Schematic of the lock-in thermography (LIT) measurement setup, a photograph of the bar-shaped $Mn_{1-a}Cr_aSb$ samples, and extraction process of thermoelectric signals. The samples form an electrical series circuit to flow charge current **I** in a zigzag manner. Under the application of an in-plane **H** with $\mu_0 H = 0.33$ T, a square-wave **I** with amplitude $I = 1$ A orthogonal to **H** was applied to the samples. The Ettingshausen effect generates a temperature gradient $\nabla T$ in the cross product direction of **I** and **H**, and results in the temperature $T$ modulations of the top surfaces oscillating at the same frequency $f$ as **I**. **b–d**, The obtained topography (**b**) and $H$-odd-dependent amplitude $A_{odd}$ (**c**) and phase $\varphi_{odd}$ (**d**) images transformed through the Fourier analysis measured at various $T$. **e,f** The $f$ dependence of $A_{odd}$ (**e**) and $\varphi_{odd}$ (**f**). **g**, The $T$ dependence of $A_{odd}$ at $f = 0$ Hz ($A_{odd}^{0\ Hz}$) normalized by the applied charge current density $j$ for all the $Mn_{1-a}Cr_aSb$ samples.

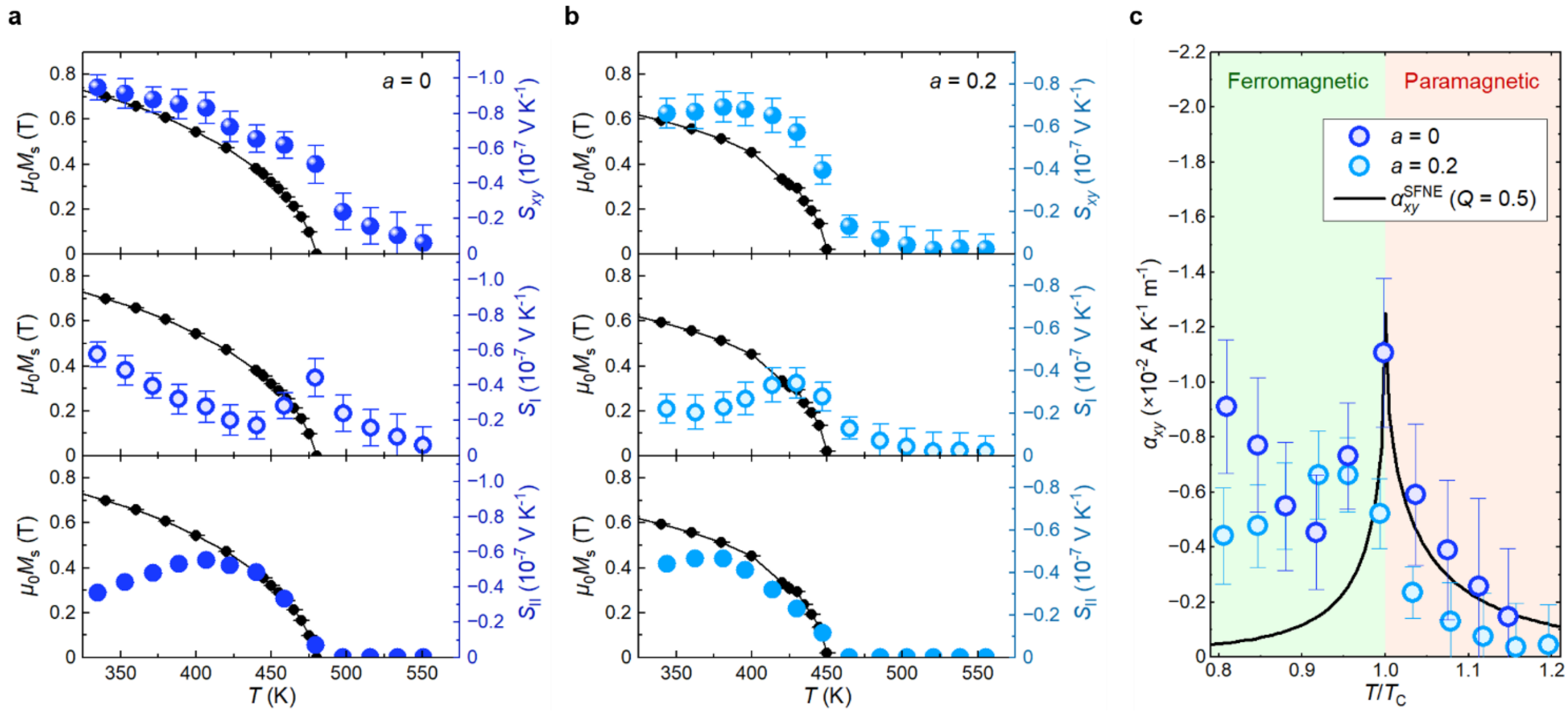

**Fig. 4 | Observation of SFNE in $Mn_{1-a}Cr_aSb$. a,b**, The $T$ dependence of $S_{xy}$ and the composed $S_{\mathrm{I}}$ and $S_{\mathrm{II}}$ terms comparing with the spontaneous magnetization $M_s$ for the $Mn_{1-a}Cr_aSb$ samples with $a = 0$ (**a**) and 0.2 (**b**) having higher $T_C$. **c**, The $T$ dependence of experimentally determined $\alpha_{xy}$ and theoretically calculated $\alpha_{xy}^{\mathrm{SFNE}}$ normalized by $T_C$ for the $Mn_{1-a}Cr_aSb$ samples with $a = 0$ and 0.2, both of which show the peak near $T/T_C = 1$.

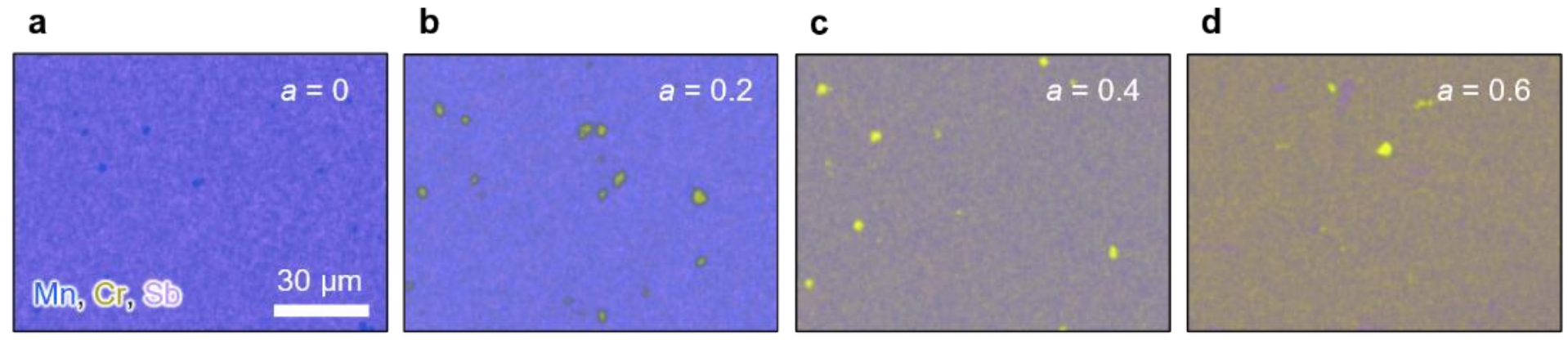

**Extended Data Fig. 1 | Microstructure observation. a–d**, SEM-EDX elemental maps of Mn (blue), Cr (yellow), and Sb (purple) obtained from the $Mn_{1-a}Cr_aSb$ samples with $a = 0$ (**a**), 0.2 (**b**), 0.4 (**c**), and 0.6 (**d**).

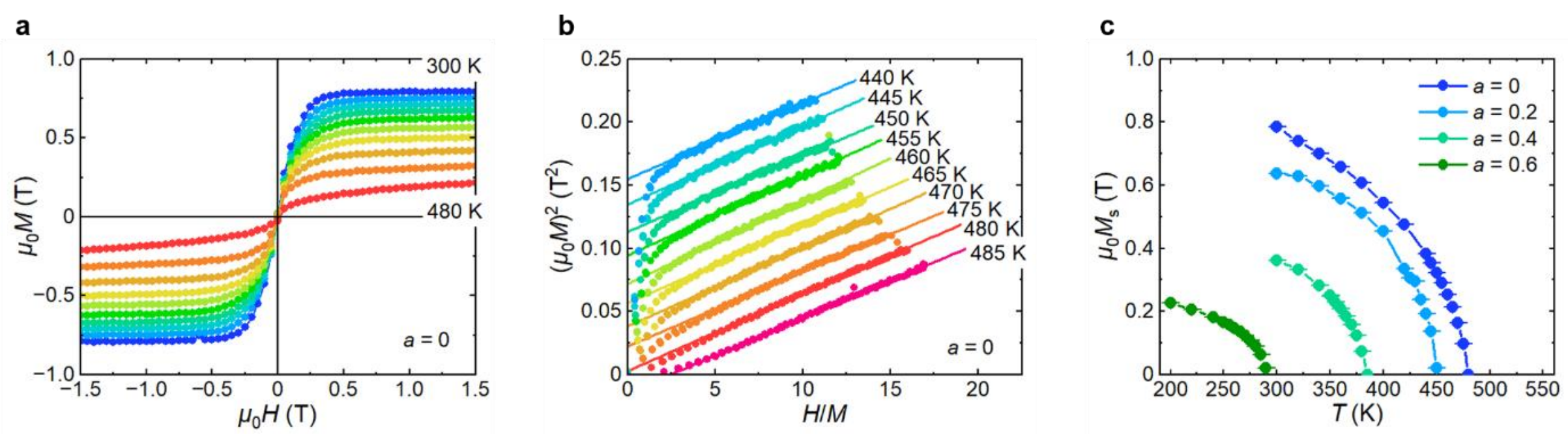

**Extended Data Fig. 2 | Temperature dependence of magnetization for $Mn_{1-a}Cr_aSb$. a,b**, The *M*–*H* curves at various *T* (**a**) and the Arrott plot (**b**) for the $Mn_{1-a}Cr_aSb$ samples with $a = 0$. **c**, The *T* dependence of the spontaneous magnetization $M_s$.

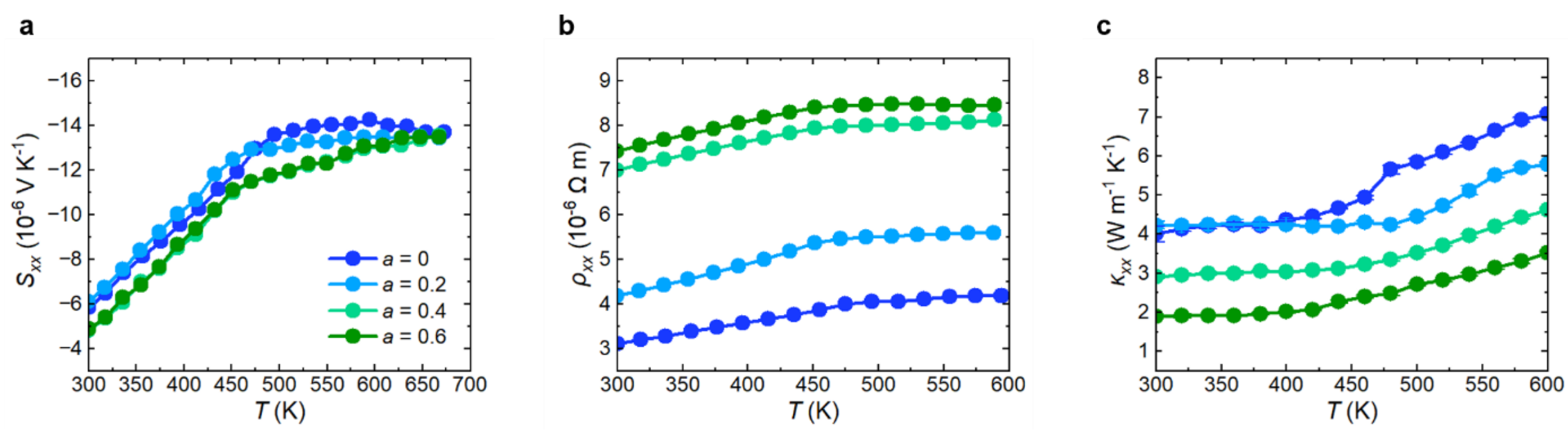

**Extended Data Fig. 3 | Temperature dependence of thermoelectric properties for $Mn_{1-a}Cr_aSb$. a–c**, The *T* dependence of the Seebeck coefficient $S_{xx}$ (**a**), electrical resistivity $\rho_{xx}$ (**b**), and thermal conductivity $\kappa_{xx}$ (**c**) for the $Mn_{1-a}Cr_aSb$ samples.

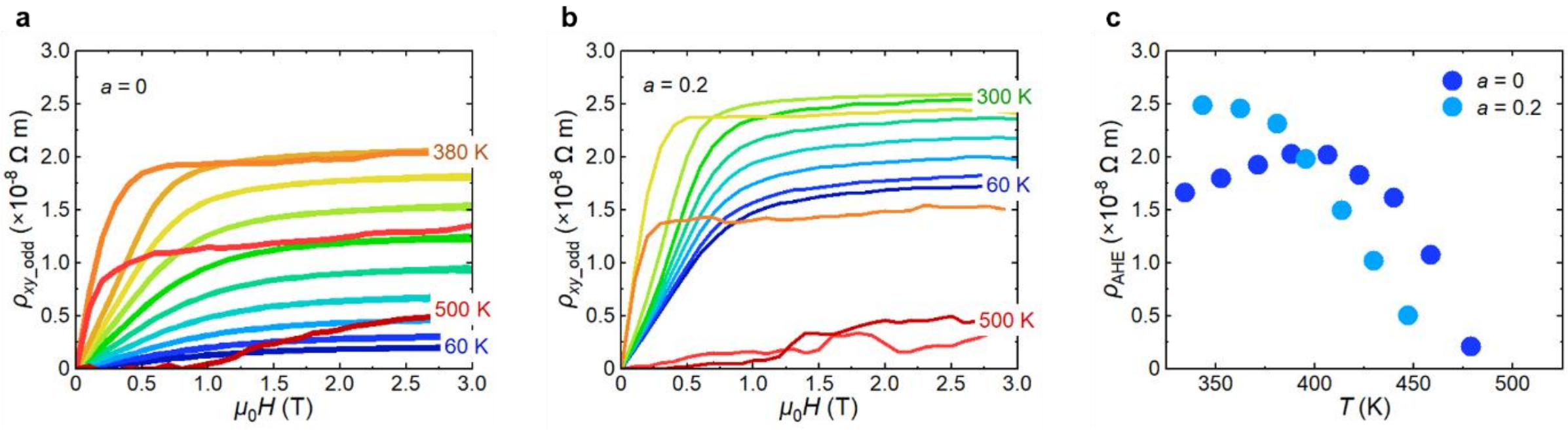

**Extended Data Fig. 4 | Temperature dependence of the anomalous Hall resistivities for $Mn_{1-a}Cr_aSb$. a–c**, The $H$ dependence of $\rho_{xy_odd}$ at various $T$ for the $Mn_{1-a}Cr_aSb$ samples with $a$ = 0 (**a**) and 0.2 (**b**) and the $T$ dependence of $\rho_{AHE}$ (**c**).

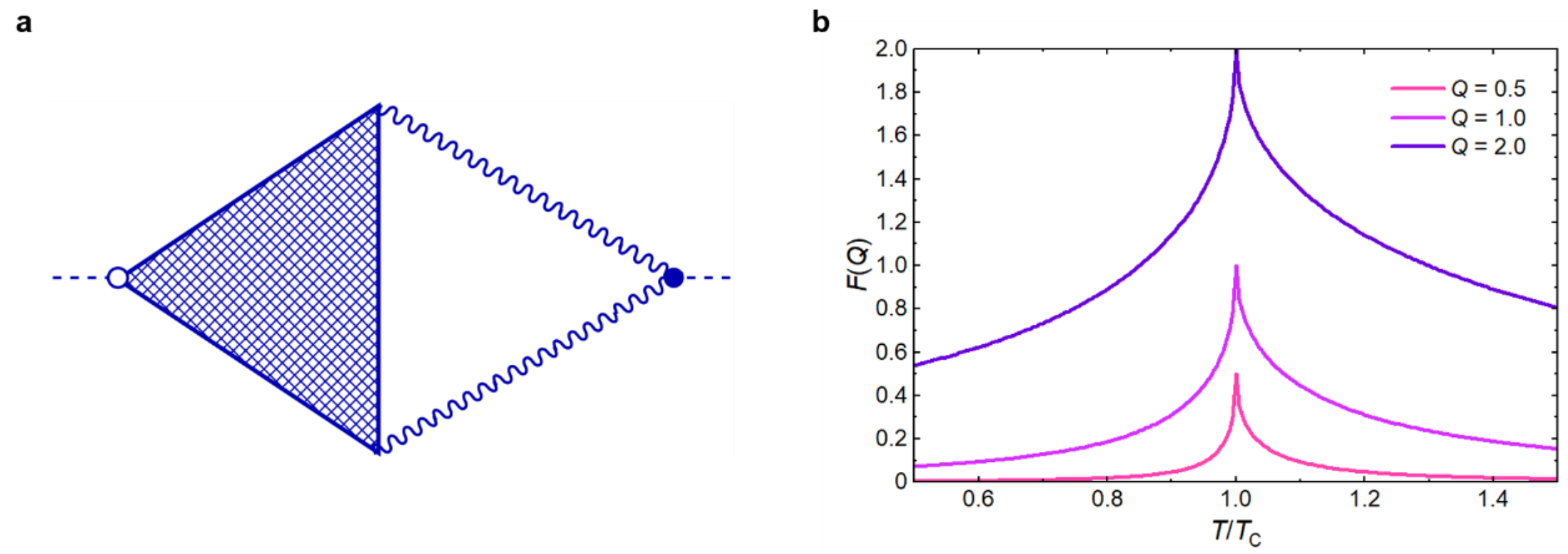

**Extended Data Fig. 5 | Semi-phenomenological modeling of SFNE. a**, Diagrammatic representation of the process for SFNE. The wavy and solid lines represent the spin fluctuation propagator and the electron Green's function, respectively. The open (filled) circle denotes the charge (heat) current vertex. The hatched triangle represents the charge current vertex renormalization by the spin-orbit interaction. **b**, $T$ dependence of $F(Q)$ for $\alpha_{xy}^{\mathrm{SFNE}}$ in Eq. (12).